\documentclass[english,preprint,nofootinbib]{revtex4}%
\usepackage[T1]{fontenc}
\usepackage[latin9]{inputenc}
\usepackage{amsmath}
\usepackage{amssymb}
\usepackage{amsfonts}
\usepackage{babel}
\usepackage{graphicx}%
\providecommand{\U}[1]{\protect\rule{.1in}{.1in}}
\makeatletter
\@ifundefined{textcolor}{}
{
\definecolor{BLACK}{gray}{0}
 \definecolor{WHITE}{gray}{1}
 \definecolor{RED}{rgb}{1,0,0}
 \definecolor{GREEN}{rgb}{0,1,0}
 \definecolor{BLUE}{rgb}{0,0,1}
 \definecolor{CYAN}{cmyk}{1,0,0,0}
 \definecolor{MAGENTA}{cmyk}{0,1,0,0}
 \definecolor{YELLOW}{cmyk}{0,0,1,0}
 }
\makeatother
\begin{document}
\title{Higher spin black hole entropy in three dimensions}
\author{Alfredo Pérez$^{1}$, David Tempo$^{1}$, Ricardo Troncoso$^{1,2}$}
\email{aperez, tempo, troncoso@cecs.cl}
\affiliation{$^{1}$Centro de Estudios Científicos (CECs), Casilla 1469, Valdivia, Chile}
\affiliation{$^{2}$Universidad Andrés Bello, Av. República 440, Santiago, Chile.}
\preprint{CECS-PHY-12/07}

\begin{abstract}
A generic formula for the entropy of three-dimensional black holes endowed
with a spin-3 field is found, which depends on the horizon area $A$ and its
spin-3 analogue $\varphi_{+}^{1/3}$, given by the reparametrization invariant
integral of the induced spin-3 field at the spacelike section of the horizon.
From this result it can be shown that the absolute value of $\varphi_{+}$ has
to be bounded from above according to $\left\vert \varphi_{+}\right\vert
^{1/3}\leq A/\sqrt{3}$. The entropy formula is constructed by requiring the
first law of thermodynamics to be fulfilled in terms of the global charges
obtained through the canonical formalism. For the static case, in the weak
spin-3 field limit, our expression for the entropy reduces to the result found
by Campoleoni, Fredenhagen, Pfenninger and Theisen, which has been recently
obtained through a different approach.

\end{abstract}
\maketitle

\section{Introduction}

The successful formulation of a consistent classical field theory that
describes interacting massless fields of spin greater than two
\cite{VV0,VV1,VV2} has recently attracted a great deal of attention, mainly
coming from different research branches within the theoretical high energy
physics community (For recent reviews see e.g., \cite{ReviewHS0,ReviewHS1,
ReviewHS3, ReviewHS4,ReviewHS5,ReviewHS6}). The three-dimensional case has
been shown to be particularly useful in order to gather more intuition about
this subject, in which spacetime and gauge symmetries become intrinsically
mixed in an unfamiliar way. Indeed, in three spacetime dimensions it is
possible to formulate a simpler theory that contains non-propagating but
interacting fields of spins two and three only, without the need of
introducing the whole tower of higher spin fields \cite{Blencowe,
BBS,Theisen-HS, AragoneDeser}. The Lagrangian is given by a Chern-Simons form
constructed out from two copies of $SL\left(  3,\mathbb{R}\right)  $, so that
the theory describes gravity with negative cosmological constant, nonminimally
coupled to an interacting spin-three field. The remarkable simplicity of this
theory further allows to find exact black hole solutions endowed with a
nontrivial spin-three field, as it has been recently reported in \cite{GK,
GKMP, CM}. Note that, since the spin-three field is nonminimally coupled to
gravity, one should not expect the standard Bekenstein-Hawking formula for the
black hole entropy to apply for this class of higher spin black holes. In
fact, in the simpler case of black holes with nonminimally coupled scalar
hair, it is known that the entropy not only depends on the area, but also on
the value of the matter field at the horizon \cite{NOS, HMTZ-2+1, ACS, MST}.
It is then natural to think that a similar effect should occur in the case of
higher spin black holes. One of the purposes of this paper is to show that
this is indeed the case. For the theory under consideration, the higher spin
black hole entropy is found to be given by%
\begin{equation}
S=\frac{A}{4G}\cos\left[  \frac{1}{3}\arcsin\left(  3^{3/2}\frac{\varphi_{+}%
}{A^{3}}\right)  \right]  \ ,\label{EntropyHS}%
\end{equation}
where $A$ and $\varphi_{+}^{1/3}$ stand for the reparametrization invariant
integrals of the induced metric and spin-3 field at the spacelike section of
the horizon, respectively; i.e., the horizon area%
\begin{equation}
A=\int_{\partial\Sigma_{+}}\left(  g_{\mu\nu}\frac{dx^{\mu}}{d\sigma}%
\frac{dx^{\nu}}{d\sigma}\right)  ^{1/2}d\sigma\ ,
\end{equation}
and its spin-three analogue%
\begin{equation}
\varphi_{+}^{1/3}:=\int_{\partial\Sigma_{+}}\left(  \varphi_{\mu\nu\rho}%
\frac{dx^{\mu}}{d\sigma}\frac{dx^{\nu}}{d\sigma}\frac{dx^{\rho}}{d\sigma
}\right)  ^{1/3}d\sigma\ .
\end{equation}

Note that the presence of the spin-3 field does not spoil the Bekenstein
bound, since the higher spin black hole entropy (\ref{EntropyHS}) is bounded
as $S\leq A/\left(  4G\right)  $. Moreover, consistency of eq.
(\ref{EntropyHS}) implies that the absolute value of $\varphi_{+}$ also has to
be bounded from above according to%
\begin{equation}
\left\vert \varphi_{+}\right\vert ^{1/3}\leq\frac{A}{\sqrt{3}}\ .
\label{BoundS-HS}%
\end{equation}
The expression for the higher spin black hole entropy in (\ref{EntropyHS}) can
be readily obtained by virtue of the first law of thermodynamics. In order to
perform this task, one short and simple path turns out to be working in the
canonical ensemble, since in this case, apart from the Hawking temperature,
one just need to know the variation of the total energy of the system. In this
way one circumvents the explicit computation of higher spin charges and their
corresponding chemical potentials that contribute\ to the work terms in the
grand canonical ensemble. As it was shown in \cite{PTTHS1}, the canonical
approach of \cite{Regge-Teitelboim} applied to the Chern-Simons theory for
$SL\left(  3,\mathbb{R}\right)  \times SL\left(  3,\mathbb{R}\right)  $ (see
also e.g. \cite{Balachandran, Banados-Q, Carlip-Q}), allows to obtain the
variation of the total energy as the surface integral that corresponds to the
variation of the total Hamiltonian, which is given by%
\begin{equation}
\delta E=\delta Q\left(  \partial_{t}\right)  =\frac{k}{2\pi}\int
_{\partial\Sigma}\left(  \left\langle A_{t}^{+}\delta A_{\theta}%
^{+}\right\rangle -\left\langle A_{t}^{-}\delta A_{\theta}^{-}\right\rangle
\right)  d\theta\ , \label{DeltaE}%
\end{equation}
where $\partial\Sigma$ stands for the boundary of the spacelike section. The
level is determined by the Newton constant and the AdS radius according to
$k=\frac{l}{4G}$, while $A_{\mu}^{\pm}$ correspond to the gauge fields
associated to both copies of $SL\left(  3,\mathbb{R}\right)  $. Here the
bracket $\left\langle \cdots\right\rangle $ is given by a quarter of the trace
in the fundamental representation (see e.g., \cite{Theisen-HS}).

Since (\ref{DeltaE}) corresponds to the variation of the energy in the
canonical ensemble, the higher spin black hole entropy (\ref{EntropyHS}) can
be found from the first law $\delta E=T\delta S$, where $T$ stands for the
Hawking temperature. In what follows\textbf{ }an explicit check of this
approach is carried out for the higher spin black hole solutions found in
\cite{GK, GKMP} and \cite{CM}\footnote{Different aspects of the thermodynamics
of these solutions have been discussed in \cite{Banados-Theisen, DFK, CLYW2}.}.

\section{Explicit examples}

\textit{(i)} The exact solution that has been found in \cite{GK, GKMP} is
described by%
\begin{equation}
A^{\pm}=g_{\pm}^{-1}a^{\pm}g_{\pm}+g_{\pm}^{-1}dg_{\pm}\ , \label{Amn}%
\end{equation}
where $g_{\pm}=g_{\pm}\left(  \rho\right)  $ correspond to suitable group
elements of each copy of $SL\left(  3,\mathbb{R}\right)  $ that depend only on
the radial coordinate, and%
\begin{align}
a^{\pm}  &  =\pm\left(  L_{\pm1}^{\pm}-\frac{2\pi}{k}\mathcal{L}L_{\mp1}^{\pm
}\mp\frac{\pi}{2k}\mathcal{W}W_{\mp2}^{\pm}\right)  dx^{\pm}\nonumber\\
&  +\mu\left(  W_{\pm2}^{\pm}-\frac{4\pi}{k}\mathcal{L}W_{0}^{\pm}+\frac
{4\pi^{2}}{k^{2}}\mathcal{L}^{2}W_{\mp2}^{\pm}\pm\frac{4\pi}{k}\mathcal{W}%
L_{\mp1}^{\pm}\right)  dx^{\mp}\ , \label{amn-GPK}%
\end{align}
where $x^{\pm}=\frac{1}{\tilde{l}}t\pm\theta$. This solution describes a
static higher spin black hole, whose metric asymptotically approaches to that
of AdS$_{3}$ of radius $\tilde{l}=l/2$, so that the asymptotic behaviour is
relaxed as compared with the one of \cite{Theisen-HS}, \cite{Henneaux-HS}. As
shown in \cite{PTTHS1}, the explicit form of $g_{\pm}=g_{\pm}\left(
\rho\right)  $, is not needed to compute the variation of energy, since
(\ref{DeltaE}) reduces to%
\begin{equation}
\delta E=\delta Q\left(  \partial_{t}\right)  =\frac{k}{2\pi}\int
_{\partial\Sigma}\left(  \left\langle a_{t}^{+}\delta a_{\theta}%
^{+}\right\rangle -\left\langle a_{t}^{-}\delta a_{\theta}^{-}\right\rangle
\right)  d\theta\ . \label{DeltaEamn}%
\end{equation}
Therefore, evaluating (\ref{DeltaEamn}) for the gauge fields $a^{\pm}$ in
(\ref{amn-GPK}), the variation of the energy was found to be given by%
\begin{equation}
\delta E=\frac{8\pi}{l}\left[  \delta\mathcal{L}-\frac{32\pi}{3k}%
\delta(\mathcal{L}^{2}\mu^{2})+\mu\delta\mathcal{W}+3\mathcal{W}\delta
\mu\right]  \ . \label{DeltaE-GPK}%
\end{equation}
As explained in \cite{GKMP}, it is useful to introduce the following change of
variables%
\begin{equation}
\frac{C-1}{C^{3/2}}=\sqrt{\frac{k}{32\pi\mathcal{L}^{3}}}\mathcal{W\ \ }%
;\ \ \gamma=\sqrt{\frac{2\pi\mathcal{L}}{k}}\mu\ , \label{CHVariables}%
\end{equation}
since the conditions that are obtained from requiring the holonomy along the
thermal cycle for the Euclidean solution to be trivial become fairly
simplified. One of these conditions allows to find the Hawking temperature
$T=\beta^{-1}$, with%
\begin{equation}
\beta=\frac{l}{4}\sqrt{\frac{\pi k}{2\mathcal{L}}}\frac{\left(  2C-3\right)
}{\left(  C-3\right)  }\left(  1-\frac{3}{4C}\right)  ^{-\frac{1}{2}}\ ,
\label{Beta-GKMP}%
\end{equation}
where in eq. (\ref{Beta-GKMP}) we have made use of the remaining
condition\footnote{The inverse Hawking temperature in (\ref{Beta-GKMP})
differs from the one in \cite{GK,GKMP} by a factor $1/2$, since here we are
using the time scale of the asymptotic region that corresponds to AdS$_{3}$
spacetime of radius $\tilde{l}=l/2$.}. The variation of the energy
(\ref{DeltaE-GPK}) then reads%
\begin{equation}
\delta E=\frac{8\pi}{l}\frac{\left(  C-3\right)  }{\left(  2C-3\right)  ^{2}%
}\left[  \left(  4C-3\right)  \delta\mathcal{L}-\frac{9}{C}\frac{\left(
2C-1\right)  }{\left(  2C-3\right)  }\mathcal{L}\delta C\right]  \ .
\label{DelteEc}%
\end{equation}
Note that the variation of the canonical energy is not an exact differential.
Nevertheless, as it has to be in the canonical ensemble, the inverse Hawking
temperature $\beta$ acts as an integrating factor such that the product
$\beta\delta E$ becomes an exact differential that defines the variation of
the entropy, i.e.,%
\begin{equation}
\delta S=\beta\delta E=\delta\left[  4\pi\sqrt{2\pi k\mathcal{L}}\left(
1-\frac{3}{2C}\right)  ^{-1}\sqrt{1-\frac{3}{4C}}\right]  \ , \label{DeltaS}%
\end{equation}
and therefore, the entropy is given by%
\begin{equation}
S=4\pi\sqrt{2\pi k\mathcal{L}}\left(  1-\frac{3}{2C}\right)  ^{-1}%
\sqrt{1-\frac{3}{4C}}\ . \label{S-GKMP}%
\end{equation}
In terms of the horizon area and the value of the purely angular component of
the spin-3 field at the horizon, which for this case read \cite{GKMP}%
\footnote{Here, the normalization of the spin-3 field agrees with \cite{CFPT},
which differs by a factor $1/6$ as compared with the one in \cite{GKMP}.}%
\begin{align}
\left.  g_{\theta\theta}\right\vert _{\rho_{+}}  &  =\left(  \frac{A}{2\pi
}\right)  ^{2}=\frac{8\pi l^{2}}{k}\frac{4C^{3}+9C^{2}-36C+27}{C\left(
2C-3\right)  ^{2}}\mathcal{L}\ ,\label{AreaHGKMP}\\
\left.  \varphi_{\theta\theta\theta}\right\vert _{\rho_{+}}  &  =\frac
{\varphi_{+}}{\left(  2\pi\right)  ^{3}}=-\left(  \frac{8\pi l^{2}\mathcal{L}%
}{kC}\right)  ^{3/2}\frac{\left(  C-1\right)  \left(  4C^{2}-3C+9\right)
}{\left(  2C-3\right)  ^{2}}\ , \label{VarphiHGKMP}%
\end{align}
one verifies that the entropy formula in eq. (\ref{EntropyHS}) reduces to
(\ref{S-GKMP}).

It is worth pointing out that the perturbative expansion of the higher spin
black hole entropy in the spin-three field, in terms of the original variables
$\mathcal{L}$ and $\mathcal{W}$ reads%
\begin{equation}
S=4\pi\sqrt{2\pi k\mathcal{L}}\left[  1+\frac{9k}{256\pi}\left(
\frac{\mathcal{W}^{2}}{\mathcal{L}^{3}}\right)  +\mathcal{O}\left(
\frac{\mathcal{W}^{2}}{\mathcal{L}^{3}}\right)  ^{2}\right]  \ ,
\label{S-Perturbative}%
\end{equation}
in agreement with the result found in \cite{CFPT} through a different
approach. The full nonperturbative expression for the entropy $S$ in eq.
(\ref{S-GKMP}) differs from the one found in \cite{GKMP}, here denoted as
$\tilde{S}$, by a factor that depends on the integration constant $C$ that
characterizes the spin-three field, i.e.,$\ S=\tilde{S}\left(  1-\frac{3}%
{2C}\right)  ^{-1}$. Further comments about the comparison with the results in
\cite{CFPT} and \cite{GKMP} are discussed below.

\bigskip

\textit{(ii) }In the case of the static higher spin black hole solution found
in \cite{CM}, whose metric approaches to that of AdS$_{3}$ of radius
$\tilde{l}=l/2$, the gauge fields can also be expressed as in (\ref{Amn}),
with $g_{\pm}=e^{\pm\rho L_{0}^{\pm}}$ and
\begin{equation}
a^{\pm}=\pm\left(  \ell_{P}L_{\pm1}^{\pm}-\mathcal{L}L_{\mp1}^{\pm}\pm\Phi
W_{0}^{\pm}\right)  dx^{\pm}+\left(  \ell_{D}W_{\pm2}^{\pm}+\mathcal{W}%
W_{\mp2}^{\pm}-QW_{0}^{\pm}\right)  dx^{\mp}\ ,
\end{equation}
with\footnote{The orientability we have chosen differs from the one in
\cite{CM}, i.e., $x^{+}\leftrightarrow x^{-}$.}%
\begin{equation}
Q\ell_{P}-2\mathcal{L}\ell_{D}=0\ \ ;\ \ Q\mathcal{L}-2\mathcal{W}\ell
_{P}=0\ .
\end{equation}
Requiring the holonomy around the thermal cycle of the Euclidean solution to
be trivial, fixes the Hawking temperature as%
\begin{equation}
T=\frac{2}{\pi l}\sqrt{\mathcal{L}\ell_{p}+4Q^{2}}\ ,
\end{equation}
and also gives the condition $\Phi=4Q$.

In \cite{PTTHS1}, by making use of the surface integral in eq. (\ref{DeltaE}),
the total energy of this higher spin black hole was found to be given by%
\begin{equation}
E=\frac{1}{G}\left[  \mathcal{L}\ell_{P}+4Q^{2}\right]  =\frac{\pi^{2}l^{2}%
}{4G}T^{2}\ , \label{E-CM}%
\end{equation}
so that by virtue of the first law $\delta E=T\delta S$, the entropy was shown
to agree with Cardy formula, i.e.,%
\begin{equation}
S=\pi l\sqrt{\frac{E}{G}}=4\pi\sqrt{\frac{c}{6}L_{0}}\ , \label{S-CM}%
\end{equation}
where, since the asymptotic form of the metric approaches to AdS$_{3}$ of
radius $\tilde{l}$, the zero modes of the Virasoro generators are given by%
\begin{equation}
L_{0}:=L_{0}^{\pm}=\frac{\tilde{l}}{2}E=\frac{l}{4}E\ ,
\end{equation}
and $c$ stands for the central charge that was shown to agree with the one of
Brown and Henneaux in \cite{Henneaux-HS, Theisen-HS, Campoleoni-HS}, i.e.,
$c=3l/(2G)$.

In this case, the value of the purely angular component of the spin-3 field at
the horizon, and the horizon area are given by
\begin{align}
\left.  \varphi_{\theta\theta\theta}\right\vert _{\rho_{+}}  &  =\frac
{\varphi_{+}}{\left(  2\pi\right)  ^{3}}=-\frac{32}{3}l^{3}Q\left(
\mathcal{L}\ell_{p}+\frac{20}{9}Q^{2}\right)  \ ,\\
\left.  g_{\theta\theta}\right\vert _{\rho_{+}}  &  =\left(  \frac{A}{2\pi
}\right)  ^{2}=4l^{2}\left(  \mathcal{L}\ell_{p}+\frac{28}{3}Q^{2}\right)  \ ,
\end{align}
respectively, from which it can seen that the entropy in eq. (\ref{S-CM}) for
this higher spin black hole is also successfully reproduced by virtue of the
generic entropy formula in (\ref{EntropyHS}).

\section{Discussion and final remarks}

It has been shown that the entropy of black holes endowed with a spin-3 field
in three dimensions can be expressed in terms of the reparametrization
invariant integrals of the pullback of the metric and the spin-3 field at the
spacelike section of the horizon, namely, the horizon area $A$, and its spin-3
analogue $\varphi_{+}^{1/3}$. The entropy formula (\ref{EntropyHS}) turns out
to be, by construction, invariant under proper gauge transformations. Indeed,
since the variation of the total energy (\ref{DeltaE}) is invariant by
definition, as well as it is the temperature, because it can be found through
requiring the holonomy along the thermal cycle to be trivial, by virtue of the
first law, $\delta S=\beta\delta E$, the entropy does.

Consistency of (\ref{EntropyHS}) also implies that the absolute value of
$\varphi_{+}^{1/3}$ becomes bounded from above by the horizon area, as in eq.
(\ref{BoundS-HS}). It is then worth pointing out that for the allowed range of
parameters for which the higher spin black holes of refs. \cite{GK, GKMP} and
\cite{CM} are defined, eq. (\ref{BoundS-HS}) turns out to be at most
saturated, which reassuringly means that the bound cannot be improved.

Note that as expressed by formula (\ref{EntropyHS}), the higher spin black
hole entropy turns out to be described by a multivalued function, so that in
order to suitably take into account the different branches of the codomain, it
is useful to parametrize the quotient $\varphi_{+}/A^{3}$ by an
\textquotedblleft angular\textquotedblright\ variable, defined as
\begin{equation}
\sin\left(  \frac{\phi}{2}\right)  :=3^{3/2}\frac{\varphi_{+}}{A^{3}}\ ,
\label{Angle}%
\end{equation}
so that the entropy reads%
\begin{equation}
S=\frac{A}{4G}\cos\left(  \frac{\phi}{6}\right)  \ . \label{Entropy-angle}%
\end{equation}
Positivity of the entropy then implies that the angle fulfills $|\phi|\leq
3\pi$. In terms of this variable, the bound in (\ref{BoundS-HS}) becomes
saturated for $|\phi|=\pi$ or $|\phi|=3\pi$, so that and in these cases the
entropy can be either $S=\frac{\sqrt{3}}{2}\frac{A}{4G}$, or $S=0$,
respectively. It is also worth highlighting that $\varphi_{+}$ may vanish even
in the presence of a nontrivial spin-3 field along spacetime, which
corresponds to the cases $\phi=0$, and $|\phi|=2\pi$. In the former case (no
winding) the entropy reduces to the one of a BTZ black hole constructed out
from identifications of the AdS$_{3}$ vacuum of radius $l$, i.e., $S=A/(4G)$,
while for the latter (winding one), the entropy is given by $S=A/(8G)$, which
also corresponds to the one of a BTZ black hole, but constructed out from the
other maximally symmetric vacuum, described by AdS$_{3}$ spacetime of radius
$\tilde{l}=l/2$.

\bigskip

In the case of static circularly symmetric higher spin black holes, since the
metric can be written in diagonal form, the horizon is located at a fixed
value of the radial coordinate $\rho=\rho_{+}$, so that the cube of the spin-3
analogue of the area becomes determined by the purely angular component of the
spin-3 field at the horizon according to $\varphi_{+}=\left(  2\pi\right)
^{3}\left.  \varphi_{\theta\theta\theta}\right\vert _{\rho_{+}}$. Hence, the
entropy (\ref{EntropyHS}) reduces to%
\begin{equation}
S=\left.  \frac{A}{4G}\cos\left[  \frac{1}{3}\arcsin\left(  3^{3/2}%
\frac{\varphi_{\theta\theta\theta}}{g_{\theta\theta}^{3/2}}\right)  \right]
\right\vert _{\rho_{+}}\ , \label{Entropy-static}%
\end{equation}
where $A=\left.  2\pi\sqrt{g_{\theta\theta}}\right\vert _{\rho_{+}}$. It is
then reassuring to verify that in the weak spin-3 field limit, formula
(\ref{Entropy-static}) expands as
\begin{equation}
S=\frac{A}{4G}\left.  \left(  1-\frac{3}{2}\left(  g^{\theta\theta}\right)
^{3}\varphi_{\theta\theta\theta}^{2}+\mathcal{O}\left(  \varphi^{4}\right)
\right)  \right\vert _{\rho_{+}}\ , \label{Entropy-pert}%
\end{equation}
in full agreement with the result found by Campoleoni, Fredenhagen, Pfenninger
and Theisen \cite{CFPT}, which has been recently obtained through a completely
different approach. Indeed, in ref. \cite{CFPT}, the action was explicitly
expressed in terms of the metric and the perturbative expansion of the spin-3
field up to quadratic order, so that the correction to the area law described
by (\ref{Entropy-pert}) was found by means of Wald's formula
\cite{Wald-Entropy}.

The discrepancy of our results for the higher spin black hole entropy, as
compared with the ones in \cite{GKMP}\footnote{For an interpretation of the
proposal in \cite{GKMP}, in terms of an underlying dual CFT in two dimensions,
see also \cite{PK, GHJ}.}, stems from a mismatch in the global charges that
becomes inherited by the entropy once computed through the first law. In fact,
as explained in \cite{PTTHS1}, since the higher spin black hole solution found
in \cite{GK, GKMP} possesses a relaxed asymptotic behavior with respect to the
one described in refs. \cite{Henneaux-HS, Theisen-HS, Campoleoni-HS}, the
global charges, and in particular, the variation of the total energy
(\ref{DeltaE}), do not depend linearly on the deviation of the fields with
respect to the reference background, so that the additional nonlinear
contributions appearing in (\ref{DeltaE-GPK}) cannot be neglected even in the
weak spin-3 field limit.

\bigskip

As an ending remark, it would be interesting to explore how the higher spin
black hole entropy formula (\ref{EntropyHS}) and the bound (\ref{BoundS-HS})
extend to the case of theories that include fields with spin greater than
three, as it is the case for the solutions discussed in \cite{PK, TAN, GGR,
CLYW1}, as well as how could they be recovered, along the lines of \cite{PK,
GHJ, H1,H2, H4,H5, H6, H7}, from a suitable notion of energy for the dual
theory at the boundary.

\acknowledgments We thank G. Barnich, A. Campoleoni, G. Compere, C.
Mart\'inez, M. Henneaux and C. Troessaert, for useful discussions and
comments. R.T. thanks the Physique th\'eorique et math\'ematique group of the
Universit\'e Libre de Bruxelles and the International Solvay Institutes for
the kind hospitality, and the Instituut voor Theoretische Fysica, Katholieke
Universiteit Leuven, for the opportunity of presenting this work. This
research has been partially supported by Fondecyt grants N${^{\circ}}$
1095098, 1121031, 3110122, 3110141, and by the Conicyt grant ACT-91:
\textquotedblleft Southern Theoretical Physics Laboratory\textquotedblright%
\ (STPLab). The Centro de Estudios Cient\'ificos (CECs) is funded by the
Chilean Government through the Centers of Excellence Base Financing Program of Conicyt.


\end{document}